\documentclass{aastex631}

\usepackage{placeins}
\usepackage{float}
\usepackage{scrextend}

\begin{document}

\title{First Light Curve Analysis of NSVS 8294044, V1023 Her, and V1397 Her Contact Binary Systems}

\author[0000-0002-0196-9732]{Atila Poro}
\altaffiliation{poroatila@gmail.com}
\affiliation{Astronomy Department of the Raderon AI Lab., BC., Burnaby, Canada}

\author{Sabrina Baudart}
\affiliation{Double Stars Committee, Société Astronomique de France, Paris, France}

\author{Mahshid Nourmohammad}
\affiliation{Independent Astrophysics Researcher, Isfahan, Iran}

\author{Zahra Sabaghpour Arani}
\affiliation{Akhtaran Astronomical Society, Isfahan Province, Aran and Bidgol, Iran}

\author{Fatemeh Farhadi}
\affiliation{Department of Physics, Faculty of Science, Zanjan University, Zanjan, Iran}

\author{Selda Ranjbar Salehian}
\affiliation{Astronomy Students’ Scientific Association, University of Tabriz, East Azerbaijan Province, Tabriz, Iran}

\author{Ahmad Sarostad}
\affiliation{Yazd Desert Night Sky Astronomy Institute, Yazd Province, Yazd, Iran}

\author{Saeideh Ranjbaryan Iri Olya}
\affiliation{Ayaz Astronomical Association, East Azerbaijan Province, Tabriz, Iran}

\author{Maryam Hadizadeh}
\affiliation{Khayyam Astronomy Association, Fars, Fasa, Iran}

\author{AmirHossein Khodaei}
\affiliation{Department of Physics, Faculty of Science, Zanjan University, Zanjan, Iran}

\begin{abstract}
The first photometric light curve investigation of the NSVS 8294044, V1023 Her, and V1397 Her binary systems is presented. We used ground-based observations for the NSVS 8294044 system and Transiting Exoplanet Survey Satellite (TESS) data for V1023 Her and V1397 Her. The primary and secondary times of minima were extracted from all the data, and by collecting the literature, a new ephemeris was computed for each system. Linear fits for the O-C diagrams were conducted using the Markov chain Monte Carlo (MCMC) method. Light curve solutions were performed using the PHysics Of Eclipsing BinariEs (PHOEBE) Python code and the MCMC approach. The systems were found to be contact binary stars based on the fillout factor and mass ratio. V1023 Her showed the O'Connell effect, and a cold starspot on the secondary component was required for the light curve solution. The absolute parameters of the system were estimated based on an empirical relationship between orbital period and mass. We presented a new $T-M$ equation based on a sample of 428 contact binary systems and found that our three target systems were in good agreement with the fit. The positions of the systems were also depicted on the $M-L$, $M-R$, $q-L_{ratio}$, and $M_{tot}-J_0$ diagrams in the logarithmic scales.
\end{abstract}


\section{Introduction}
The degree of separation between the components describes different kinds of eclipsing binaries, which may be identified by light curve shapes and include detached, semi-detached, contact, and overcontact systems. Contact binary stars are referred to as W UMa-type systems that have some interesting physical properties (\citealt{2014ApJ...790..157D}). 

Stars of the W UMa systems have filled their Roche lobes (\citealt{1971ARA&A...9..183P}) and both of the components share a common convective envelope between the inner and outer contact surfaces (\citealt{1968ApJ...151.1123L}). The stars in these kinds of systems share mass and energy, which causes their surface temperatures to be close to one another and show light curves with almost similar eclipse depths. Furthermore, W UMa stars consist of late spectral type main sequence stars (\citealt{2012AJ....143...99T}).

The orbital period of contact binary systems is short and most of them fall between 0.2 and 0.6 days (\citealt{2003MNRAS.342.1260Q}, \citealt{2018PASJ...70...90K}, \citealt{2021ApJS..254...10L}). Studying contact systems provides an opportunity to use orbital period variations impacted by system physical processes to gain a deeper comprehension of star evolution (\citealt{2021AJ....162...13L}). Therefore, new observations can always be useful to investigate these variations.
\\
\\
This study includes the first light curve analysis of three binary systems NSVS 8294044 (R.A. $19^h$ $14^m$ $41.0475^s$, Dec. {$+37^\circ$ $49'$ $46.0465"$} (J2000)), V1023 Her (R.A. $15^h$ $58^m$ $25.2669^s$, Dec. {$+49^\circ$ $26'$ $51.0013"$} (J2000)), and V1397 Her (R.A. $17^h$ $07^m$ $16.8455^s$, Dec. {$+17^\circ$ $25'$ $36.6323"$} (J2000)). All three target systems have been introduced as contact systems in the ASAS\footnote{The All Sky Automated Survey (ASAS), \url{http://www.astrouw.edu.pl/asas/}} and VSX\footnote{The International Variable Star Index (VSX), \url{https://www.aavso.org/vsx/}} catalogs.
NSVS 8294044, V1023 Her, and V1397 Her's orbital periods are near to each other as reported at 0.3779846, 0.3222336, and 0.387745 days in the VSX database, respectively. Also, the apparent magnitude of the NSVS 8294044 is $V_{max}=13.51^{mag}$, V1023 Her is $V_{max}=11.93^{mag}$, and V1397 Her is $V_{max}=10.16^{mag}$, based on the VSX database.
\\
\\
The structure of this paper is as follows: Section 2 is about ground- and space-based observations and data reduction; Section 3 is related to determining new ephemeris for each system; Section 4 explains the light curve analysis; Section 5 describes the estimation of absolute parameters; Finally, the discussion and conclusion are in Section 6.

\vspace{1.5cm}
\section{Observation and Data Reduction}
We observed NSVS 8294044 in August and September 2023. The observations were carried out in a private observatory in Toulon, France, at a longitude of $05^{\circ}$ $54'$ $35"$ E and a latitude of $43^{\circ}$ $8'$ $59"$ N, and an altitude of 68 meters above mean sea level.
In the observations, we used an Apochromatic Refractor TS Optics with a 102mm aperture, a ZWO ASI 1600MM CCD, and a $V$ filter. The binning of the images was $1\times1$, with an 80-second exposure time, and the average temperature of the CCD was $-15^{\circ}$C.
Using the Siril 1.2.0-rc2 program and the bias, dark, and flat fields image, the basic data reduction was done.
We used GSC 02665-00244 (R.A. $19^h$ $17^m$ $21.4390^s$, Dec. {$+37^\circ$ $11'$ $51.0482"$} (J2000)) as a comparison star, and GSC 02665-01066 (R.A. $19^h$ $17^m$ $11.7567^s$, Dec. {$+37^\circ$ $13'$ $32.9641"$} (J2000)) and Gaia DR2 2051151521582207232 (R.A. $19^h$ $17^m$ $06.8187^s$, Dec. {$+37^\circ$ $04'$ $22.3909"$} (J2000)) as check stars.
According to our observations, the light curve's maximum apparent magnitude was obtained to be $V_{max}=13.24(11)^{mag}$.
\\
\\
TESS data were used in this investigation for V1023 Her (TIC 310170498) and V1397 Her (TIC 143100813) binary systems.
TESS observed V1023 Her in sectors 23, 24, 50, and 51. There were also observational data for the V1397 Her system in sectors 12 and 13. TESS employed a 120-second exposure time for the observation process of these sectors.
The data is available at the Mikulski Space Telescope Archive (MAST)\footnote{\url{https://mast.stsci.edu/portal/Mashup/Clients/Mast/Portal.htmL}}.

\vspace{1.5cm}
\section{New Ephemeris}
We extracted the primary and secondary times of minima from ground-based and TESS light curves. Gaussian distribution was used to determine these mid-eclipse times. The amount of uncertainty was computed using the MCMC method.
The Barycentric Julian Date and Barycentric Dynamic Time ($BJD_{TDB}$) are used to present all minima.
The minima for three target systems that were collected from the literature and extracted in this study are listed in Table \ref{tab1}. The extracted TESS mid-eclipse times for the V1023 Her and V1397 Her systems are listed in the appendix tables.

Each epoch and O-C values were computed using reference ephemeris. We chose $t_0=2458423.7230$ and $P_{ref}=0.3779833$ from the ASAS-SN catalog for the NSVS 8294044 system as a reference ephemeris. V1023 Her's $t_0=2459636.8875$ came from \cite{2022OEJV..234....1N} study and $P_{ref}=0.3222341(87)$ from the WISE catalog of periodic variable stars. The reference ephemeris of V1397 Her $t_0=2453833.6397$ and $P_{ref}=0.387745$ were from the \cite{2009IBVS.5894....1D} study.

\begin{figure*}
\begin{center}
\includegraphics[width=\textwidth]{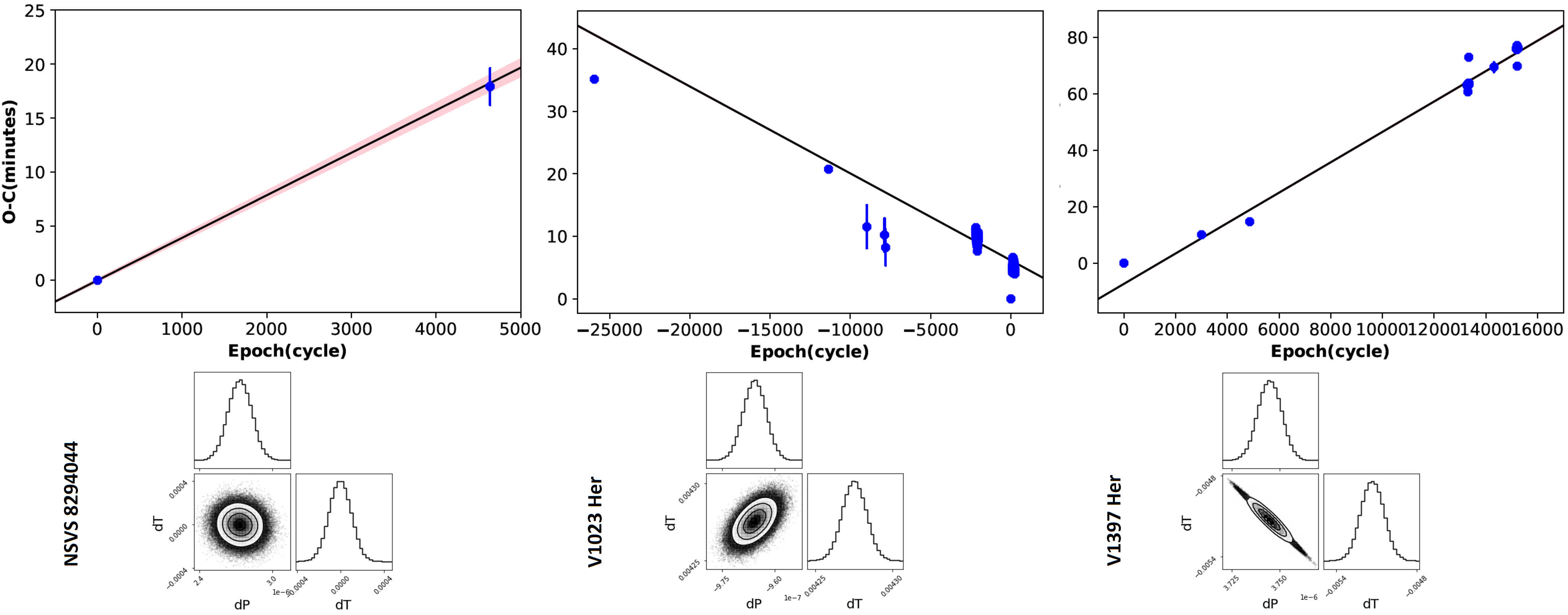}
\caption{The O-C diagrams of three eclipsing binaries with linear fits and corner plots. The shaded regions show the model parameters' 68th percentile values.}
\label{Fig1}
\end{center}
\end{figure*}

It is more appropriate to use a linear fit for O-C diagrams considering our three binary systems have few observations and minimum times. O-C diagrams of the systems are presented in Figure \ref{Fig1} with their corner plots of the posterior distribution based on the MCMC sampling. To determine new ephemeris for each system, we applied 20 walkers and 20000 iterations for each walker, with a 2000 burn-in period in the MCMC process. So, we used the PyMC3 package to execute the MCMC sampling (\citealt{2016ascl.soft10016S}). New ephemeris for each target system is presented in Equations 1 to 3:

\begin{equation}\label{eq1}
\left\{\begin{array}{l}
NSVS\ 8294044:\\
Min.I(BJD_TDB)=2458423.72300(10)+0.377986034(9)\times E\\
\end{array}\right.
\end{equation}

\begin{equation}\label{eq2}
\left\{\begin{array}{l}
V1023\ Her:\\
Min.I(BJD_TDB)=2459636.89178(1)+0.322233134(3)\times E\\
\end{array}\right.
\end{equation}

\begin{equation}\label{eq3}
\left\{\begin{array}{l}
V1397\ Her:\\
Min.I(BJD_TDB)=2453833.63457(8)+0.387748745(6)\times E\\
\end{array}\right.
\end{equation}

\begin{table*}
\caption{The times of minima extracted in this study and collected from the literature were observed with CCD.}
\centering
\begin{center}
\footnotesize
\begin{tabular}{c c c c c c}
\hline
\hline
System & Min.($BJD_{TDB}$) & Error & Epoch & O-C & Reference\\
\hline
NSVS 8294044	&	2458423.7230	&		&	0	&	0	&	ASAS-SN	\\
	&	2460176.4440	&	0.0012	&	4637	&	0.0124569	&	This study	\\
	&	2460191.3750	&	0.0022	&	4676.5	&	0.01314655	&	This study	\\
\hline
V1023 Her	&	2451274.6148	&		&	-25951	&	0.0244	&	ROTSE- Anton Paschke	\\
	&	2455976.6448	&		&	-11359	&	0.0144	&	VSX	\\
	&	2456687.9726	&	0.0002	&	-9151.5	&	0.0105	&	\cite{2015IBVS.6131....1N}	\\
	&	2456745.4889	&	0.0024	&	-8973	&	0.0080	&	\cite{2015IBVS.6149....1H}	\\
	&	2457100.4327	&	0.0029	&	-7871.5	&	0.0109	&	\cite{2016IBVS.6157....1H}	\\
	&	2457100.5900	&	0.0019	&	-7871	&	0.0071	&	\cite{2016IBVS.6157....1H}	\\
	&	2457122.8228	&	0.0020	&	-7802	&	0.0057	&	\cite{2016IBVS.6164....1N}	\\
	&	2459636.8875	&		&	0	&	0	&	\cite{2022OEJV..234....1N}	\\
\hline
V1397 Her	&	2453833.6397	&		&	0	&	0	&	\cite{2009IBVS.5894....1D}	\\
	&	2454623.6747	&		&	2037.5	&	0.0046	&	VSX	\\
	&	2454998.8205	&	0.0007	&	3005	&	0.0070	&	\cite{2009IBVS.5894....1D}	\\
	&	2455720.8049	&	0.0003	&	4867	&	0.0102	&	\cite{2011IBVS.5992.1D}	\\
	&	2456054.8507	&	0.0005	&	5728.5	&	0.0137	&	\cite{2012IBVS.6029....1D}	\\
	&	2458626.0127	&		&	12359.5	&	0.0386	&	\cite{kazuo2020visual}	\\
	&	2458991.4659	&	0.0008	&	13302	&	0.0422	&	\cite{pagel2021bav}	\\
	&	2459383.4822	&	0.0013	&	14313	&	0.0483	&	\cite{pagel2022bav}	\\
\hline
\hline
\end{tabular}
\end{center}
\label{tab1}
\end{table*}

\vspace{1.5cm}
\section{Light Curve Analysis}
We utilized the PHOEBE Python code version 2.4.9 and the MCMC approach to model the light curves of the NSVS 8294044, V1023 Her, and V1397 Her binary systems (\citealt{2005ApJ...628..426P}, \citealt{2016ApJS..227...29P}, \citealt{2020ApJS..250...34C}). These three target systems were analyzed for the first time.
We selected contact mode for the light curve solutions based on the appearance of the light curve and short orbital period. The gravity-darkening coefficients and the bolometric albedo were assumed to be $g_1=g_2=0.32$ (\citealt{1967ZA.....65...89L}) and $A_1=A_2=0.5$ (\citealt{1969AcA....19..245R}), respectively. The limb darkening coefficients were included in PHOEBE as a free parameter, and the \cite{2004AA...419..725C} study was used to model the stellar atmosphere.

We applied the effective temperature reported from Gaia DR3 to the hotter star. These temperature values are not fixed and will be free parameters during the MCMC process. The effective temperature ratio was used to estimate the initial temperature of another component.
Additionally, we checked and compared the effective temperatures of these systems from TESS through the MAST portal\footnote{\url{https://mast.stsci.edu/portal/Mashup/Clients/Mast/Portal.htmL}}. The TESS input v8.2 reported temperatures 5937(115), 5088(126), and 6440(113) for NSVS 8294044, V1023 Her, and V1397 Her, respectively. These values were close to those reported by Gaia DR3.

\begin{figure*}
\begin{center}
\includegraphics[scale=0.18]{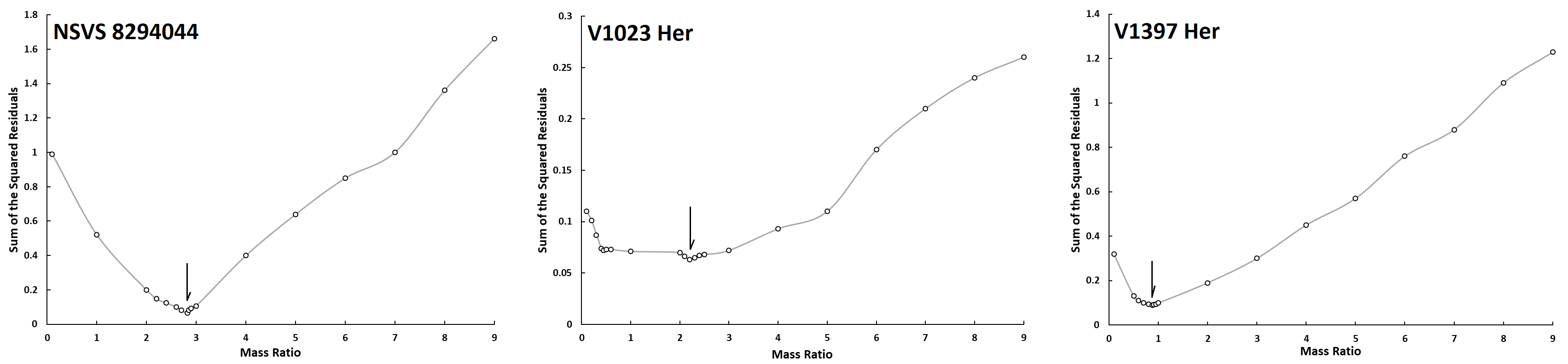}
\caption{Sum of the squared residuals as a function of the mass ratio.}
\label{Fig2}
\end{center}
\end{figure*}

The $q$-search method was used to estimate the mass ratio of each binary system, taking into account that only photometric data was available. We performed a $q$-search on each of the target systems using a range of 0.1 to 9. We found the minimum sum of the squared residuals for each $q$-search. We used the results as initial values and tried to fit a good synthetic light curve to the observational data. Figure \ref{Fig2} displays the $q$-search results for each of the three systems.

V1023 Her system's light curve analysis required a cold starspot on the secondary component due to the difference in the light curve maxima. Contact systems are known for their magnetic activity, and O'Connell's effect describes it (\citealt{1951MNRAS.111..642O}).

We employed PHOEBE's optimization tool to improve the output of light curve solutions and obtain entitled results for beginning the MCMC process. The five main parameters ($T_1$, $T_2$, $q$, $f$, $l_1$) were then processed using the MCMC approach, and the values and uncertainty were obtained. We employed 96 walkers and 1000 iterations for each walker. These walkers were initially positioned from a Gaussian distribution based on our initial parameter estimations, and their width was adjusted based on how sensitive the light curves were to various parameter values.

\begin{figure*}
\begin{center}
\includegraphics[scale=0.22]{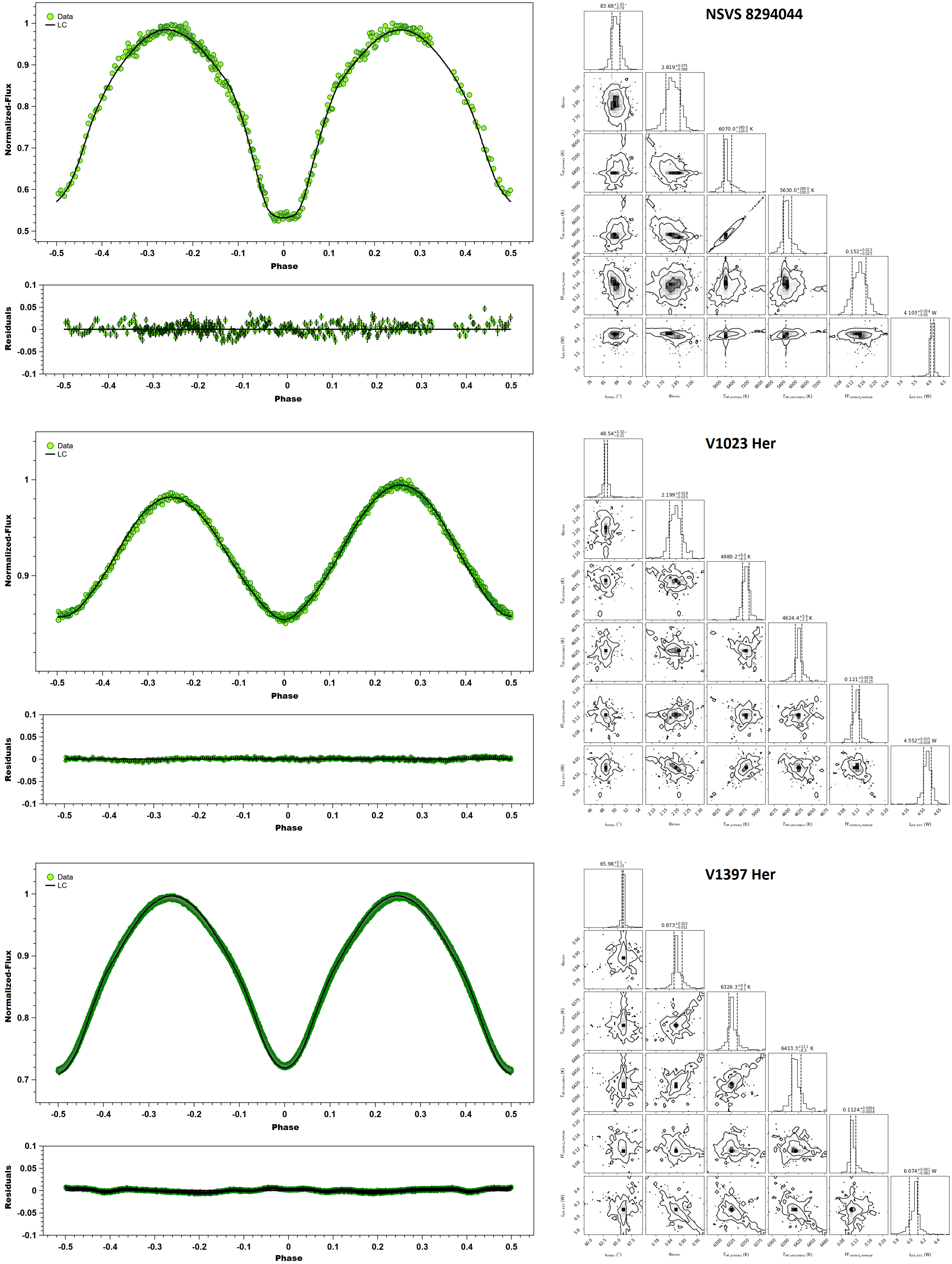}
\caption{Left: The observed light curves of the systems (green dots) and the synthetic light curves were obtained from the light curve solutions (black curves). Right: The corner plots of the three systems were determined by MCMC modeling.}
\label{Fig3}
\end{center}
\end{figure*}

\begin{figure*}
\begin{center}
\includegraphics[width=\textwidth]{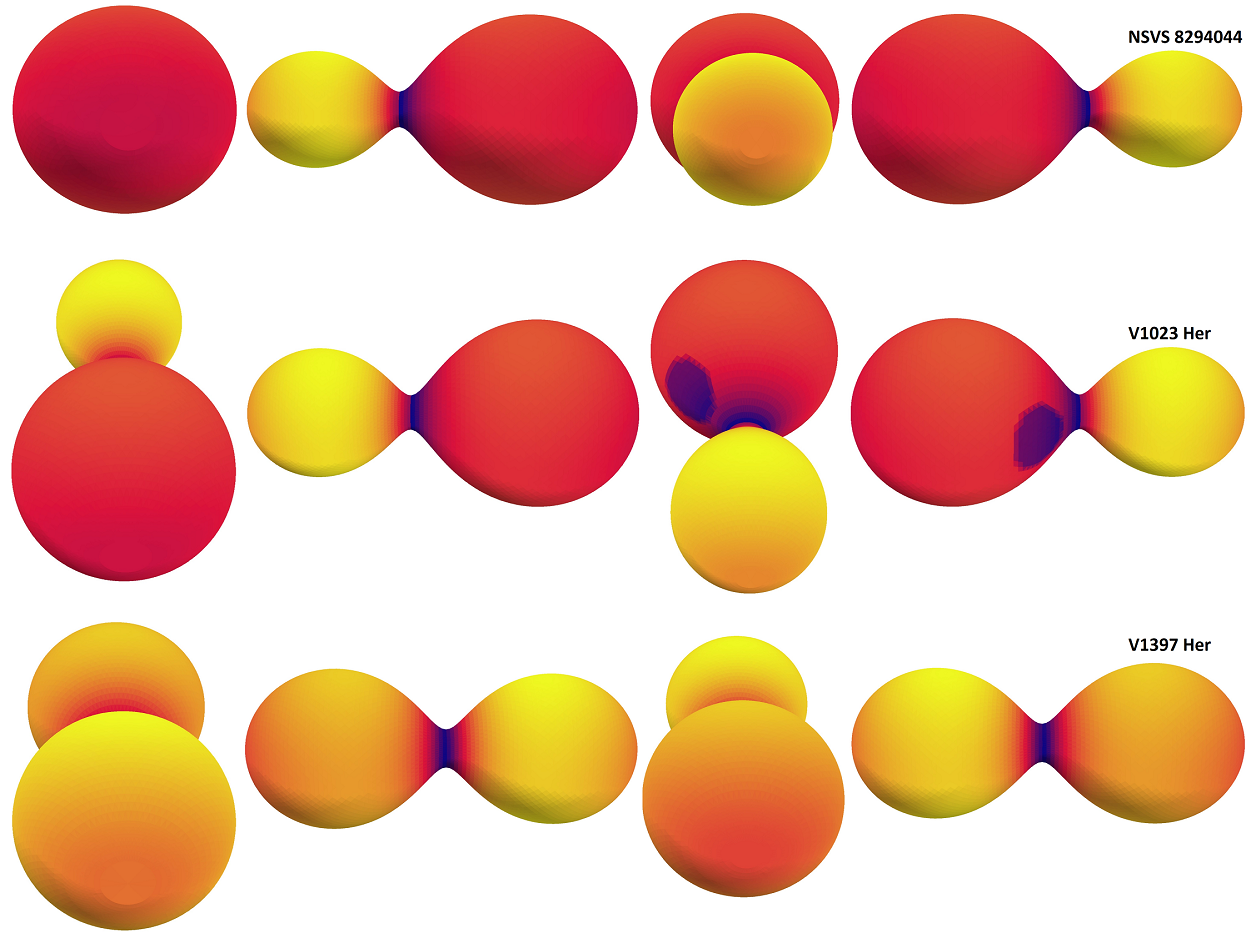}
\caption{Geometric structure of the NSVS 8294044, V1023 Her, and V1397 Her systems based on their light curve solutions in 0, 0.25, 0.5, and 0.75 phases.}
\label{Fig4}
\end{center}
\end{figure*}

The results of the light curve solutions of the NSVS 8294044, V1023 Her, and V1397 Her binary systems are listed in Table \ref{tab2}. The observed and synthetic light curves of the binary systems, along with the corner plots, are displayed in Figure \ref{Fig3}. Furthermore, the geometric structures of the systems are shown in Figure \ref{Fig4}.

\begin{table*}
\caption{Photometric light curve solutions of the systems.}
\centering
\begin{center}
\footnotesize
\begin{tabular}{c c c c}
 \hline
 \hline
Parameter & NSVS 8294044 & V1023 Her & V1397 Her\\
\hline
$T_{1}$ (K) & $6070_{\rm-(120)}^{+(320)}$ & $4980_{\rm-(6)}^{+(6)}$ & $6326_{\rm-(6)}^{+(10)}$\\
\\
$T_{2}$ (K) & $5630_{\rm-(100)}^{+(290)}$ & $4624_{\rm-(7)}^{+(6)}$ & $6413_{\rm-(8)}^{+(13)}$\\
\\
$q=M_2/M_1$ &  $2.819_{\rm-(69)}^{+(75)}$ & $2.199_{\rm-(25)}^{+(28)}$ & $0.873_{\rm-(11)}^{+(23)}$\\
\\
$i^{\circ}$ &  $83.68_{\rm-(79)}^{+(99)}$ & $48.54_{\rm-(21)}^{+(32)}$ & $65.98_{\rm-(15)}^{+(10)}$\\
\\
$f$ &  $0.152_{\rm-(25)}^{+(23)}$ & $0.121_{\rm-(13)}^{+(8)}$ & $0.112_{\rm-(6)}^{+(9)}$\\
\\
$\Omega_1=\Omega_2$ &  6.282(41) & 5.458(37) & 3.486(24)\\
\\
$l_1/l_{tot}$ & $0.359_{\rm-(3)}^{+(3)}$ & $0.406_{\rm-(1)}^{+(1)}$ & $0.520_{\rm-(4)}^{+(1)}$\\
\\
$l_2/l_{tot}$ & 0.641(3) & 0.594(1) & 0.480(2)\\
\\
$r_{1(mean)}$ & 0.303(2) & 0.322(2) & 0.402(3)\\
\\
$r_{2(mean)}$ & 0.491(2) & 0.459(2) & 0.378(3)\\
\\
Phase shift & 0.033(1) & -0.012(1) & 0.003(1)\\
\hline
Col.(deg) &  & 107 &\\
Long.(deg) &  & 320 &\\
Rad.(deg) &  & 16 &\\
$T_{spot}/T_{star}$ &  & 0.92 &\\
Component &  & Secondary &\\
\hline
\hline
\end{tabular}
\end{center}
\label{tab2}
\end{table*}

\vspace{1.5cm}
\section{Absolute Parameters}
We used empirical parameter relationships between mass and orbital period for estimating absolute parameters (\citealt{PORO2024NewAstro(b)}). So, the following equation was chosen from the \cite{2021ApJS..254...10L} study result of a large sample:

\begin{equation}\label{eq4}
M_1=(2.94\pm 0.21)P+(0.16\pm 0.08)
\end{equation}

Was considered on $q'=1/q$ in the \cite{2021ApJS..254...10L} sample and points mentioned in the \cite{2024RAA....24a5002P(c)} study. Therefore, we set the outcome's mass from Equation \ref{eq4} for the more massive star in the systems.
Using the mass ratio from the light curve solution, we obtained the mass of the other companion star. Then, we computed $a(R_\odot)$ by employing Kepler's third law. We estimated the radius of the stars of the system from the equation $R=a\times r(mean)$. The known equation $L=4\pi R^2 \sigma T^4$ is also used to determine each star's luminosity while having its effective temperature and radius. Based on Pogson’s relation (\citealt{1856MNRAS..17...12P}), $M_{bol1,2}$ can be calculated by $M_{bol}-M_{bol\odot}=-2.5log(L/L_{\odot})$, which $M_{bol\odot}$ is taken as 4.73 magnitude (\citealt{2010AJ....140.1158T}). We extracted the Bolometric Correction (BC) from the \cite{1996ApJ...469..355F} study tables and utilized the well-known equation $M_V=M_{bol}-BC$ to compute $M_{V1,2}$. The surface gravity ($g$) of the stars was calculated with the equation $g=G_{\odot}(M/R^2)$.

Estimated absolute parameters for the target systems are presented in Table \ref{tab3}. The error values of Equation \ref{eq4} and the parameters involved were used to estimate the absolute parameter uncertainty.

\begin{table*}
\caption{The absolute parameters estimation.}
\centering
\begin{center}
\footnotesize
\begin{tabular}{c | c c | c c | c c}
 \hline
 \hline
& \multicolumn{2}{c|}{NSVS 8294044} & \multicolumn{2}{c|}{V1023 Her} & \multicolumn{2}{c}{V1397 Her}\\
Parameter & Primary star & Secondary star & Primary star & Secondary star & Primary star & Secondary star\\
\hline
$M(M_\odot)$ & 0.451(44) & 1.271(160) & 0.503(61) & 1.107(148) & 1.300(161) & 1.135(165)\\
$R(R_\odot)$ & 0.799(189) & 1.295(303) & 0.746(170) & 1.064(240) &  1.210(216) & 1.138(204)\\
$L(L_\odot)$ & 0.781(512) & 1.518(962) & 0.309(159) & 0.467(239) & 2.114(836) & 1.974(788)\\
$M_{bol}$(mag.) & 4.998(547) & 4.277(533) & 6.005(451) & 5.558(448) & 3.917(362) & 3.992(365)\\
$M_{V}$(mag.) & 5.034(547) & 4.386(533) & 6.320(451) & 6.074(448) & 3.925(362) & 3.993(365)\\
$log(g)(cgs)$ & 4.287(144) & 4.318(131) & 4.394(129) & 4.428(122) & 4.386(92) & 4.381(84)\\
$a(R_\odot)$ & \multicolumn{2}{c|}{2.637(604)} & \multicolumn{2}{c|}{2.318(511)} & \multicolumn{2}{c}{3.010(511)} \\
\hline
$logJ_0$ & \multicolumn{2}{c|}{51.632(76)} & \multicolumn{2}{c|}{051.606(86)} & \multicolumn{2}{c}{51.996(91)} \\
$BC$ & -0.036 & -0.109 & -0.315 & -0.516 & -0.008 & -0.001\\
\hline
\hline
\end{tabular}
\end{center}
\label{tab3}
\end{table*}

\vspace{1.5cm}
\section{Discussion and Conclusion}
We present the first light curve analysis of the NSVS 8294044, V1023 Her, and V1397 Her binary systems. Ground-based observations were used for NSVS 8294044, and TESS observations were used for V1023 Her and V1397 Her systems. The discussion and conclusion are presented as follows:

1. We extracted times of minima from ground- and space-based photometric data. Thus, by adding the literature, new ephemeris for each system were obtained. A linear fit was the better option for the target systems on the O-C diagram. The O-C diagram of V1023 Her has a decreasing trend based on the linear fit, whereas systems NSVS 8294044 and V1397 Her display an increasing trend (Figure \ref{Fig1}).

2. Light curve analysis of each target binary system was performed and V1023 Her required a cold starspot. The results of light curve solutions indicate that the effective temperature range of the stars is between about 4600 K and 6400 K. Star1 is hotter than star2 except for system V1397 Her. The temperature difference of the components for NSVS 8294044 is 440 K, for V1023 Her is 356 K, and for V1397 Her is 87 K. According to the temperature of the stars and the \cite{2000asqu.book.....C} and \cite{2018MNRAS.479.5491E} studies, the spectral type of the stars can be recognized: Star1=G0 and star2=G7 for NSVS 8294044; Star1=K1 and star2=K3 for V1023 Her; Star1=F6 and star2=F5 for V1397 Her.

\begin{figure*}
\begin{center}
\includegraphics[width=\textwidth]{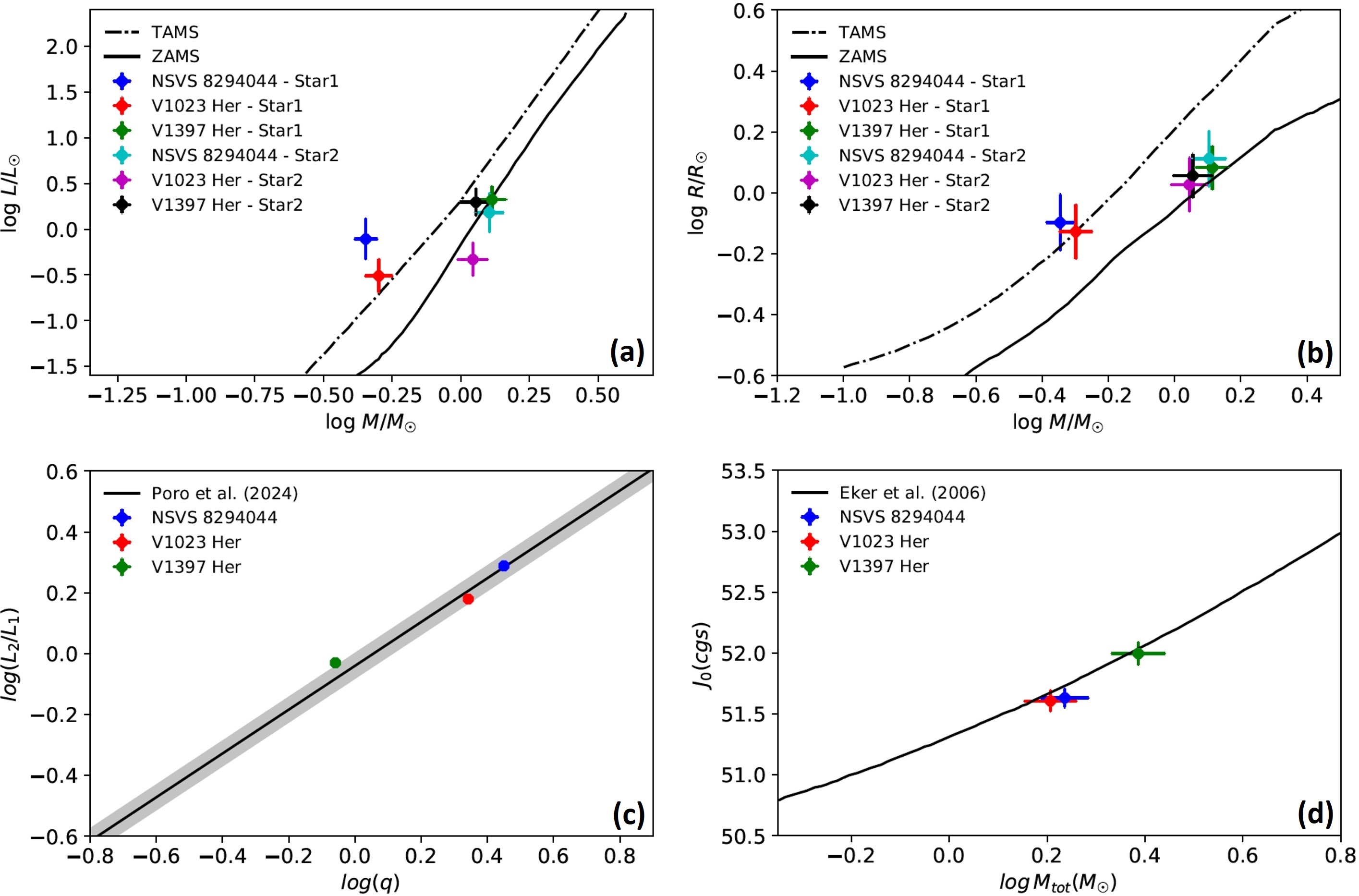}
\caption{The $logM-logL$, $logM-logR$, $logq-logL_{ratio}$, and $logM_{tot}-logJ_0$ diagrams.}
\label{Fig5}
\end{center}
\end{figure*}

3. We used an empirical parameter relationship and orbital periods for estimating the mass of each star. According to the absolute parameters, we showed the stars' positions on the Mass-Luminosity ($M-L$) and Mass-Radius ($M-R$) diagrams with the Zero-Age Main Sequence (ZAMS) and the Terminal-Age Main Sequence (TAMS) (Figure \ref{Fig5}a,b). The position of the systems on the $q-L_{ratio}$ relationship provided by \cite{2024RAA....24a5002P(c)} are also displayed in Figure \ref{Fig5}c, with which they are in good agreement with the model.

4. We calculated the orbital angular momentum ($J_0$) of the systems based on the equation presented by the \cite{2006MNRAS.373.1483E} study:

\begin{equation}\label{eq5}
J_0=\frac{q}{(1+q)^2} \sqrt[3] {\frac{G^2}{2\pi}M^5P}
\end{equation}

The results are listed in Table \ref{tab3}. Furthermore, as can be viewed in Figure \ref{Fig5}d, the $logM_{tot}-logJ_0$ diagram illustrates the systems' positions and indicates that all of them are located in the region of contact binary systems. The parabolic fit in Figure \ref{Fig5}d of the border between the two detached and contact binaries comes from the \cite{2006MNRAS.373.1483E} study.

5. The relationship between parameters in binary systems is one of the topics of interest in studies. However, most of the investigations on the mass-temperature ($M-T$)relationship have focused on detached and semi-detached binary systems (e.g. \citealt{1967AcA....17....1P}, \citealt{1988BAICz..39..329H}, \citealt{2002ARep...46..233K}, \citealt{2007MNRAS.382.1073M}, \citealt{2013ApJ...776...87S}, \citealt{2018MNRAS.479.5491E}). In the \cite{2005ApJ...629.1055Y} study, the $logM-logT$ diagram for contact binaries displays a sample of A- and W-type systems with a ZAMS line. Additionally, the $M-T$ diagram for a limited number of contact systems in the \cite{2019AJ....158..186K} study indicates a weak relationship for this sample given the large dispersion that appears.

The relationship between $P-T_1-M_1$ is presented using a machine learning model in another work by \cite{2024PASP..136b4201P(a)}. They employed 134 contact systems that were analyzed using spectroscopic data. This model is presented using the Artificial Neural Network (ANN) and $M_1$ is the more massive component of the system. 
Based on the ANN model, more massive stars 1.20(8), 0.99(9), and 1.27(6) have been estimated for systems NSVS 8294044, V1023 Her, and V1397 Her, respectively; these values are in the uncertainty range and good in agreement with the results of this study.

\begin{figure*}
\begin{center}
\includegraphics[scale=0.55]{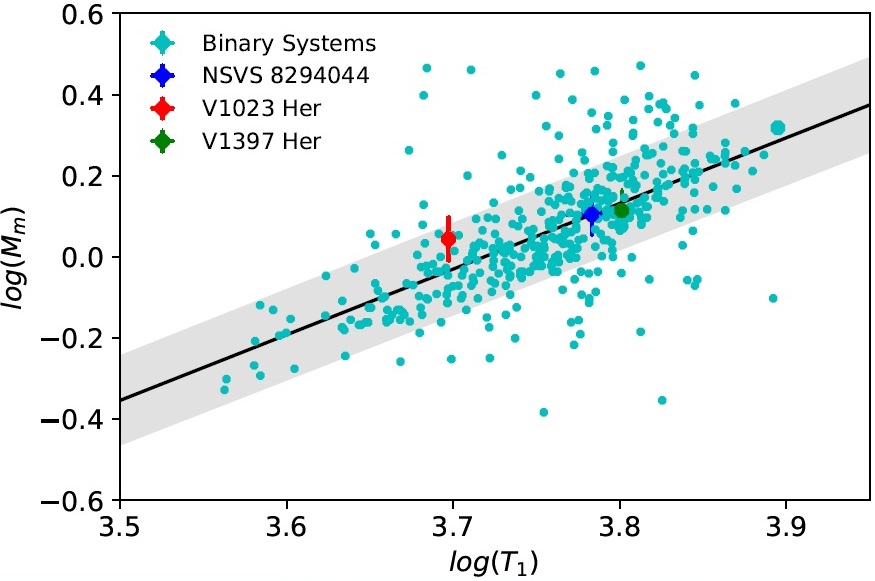}
\caption{Diagram of the effective temperature ($T_1$) and more massive component ($M_m$) of the contact binary systems.}
\label{Fig6}
\end{center}
\end{figure*}

We provided the $T_1-M_m$ diagram with a linear fit using 428 contact systems from the sample of the \cite{2021ApJS..254...10L} study (Equation \ref{eq6}). We named the more massive star $M_m$. Figure \ref{Fig6} shows the positions of our three target systems are in good agreement with this fit and the other stars in the sample.

\begin{equation}\label{eq6}
logM_m=(1.6185\pm0.0150)\times logT_1+(-6.0186\pm0.0562)
\end{equation}

6. The first light curve study of binary stars is important to create larger samples for deeper parameter investigations of these types of systems.
This investigation presented the new ephemeris and the first light curve analysis of the target binary systems. We concluded that NSVS 8294044, V1023 Her, and V1397 Her are contact binary systems based on the mass ratio, fillout factor, inclination, absolute parameters, and the $logM_{tot}-logJ_0$ diagram.

\clearpage
\section*{Data Availability}
Ground-based data will be made available on request.

\vspace{1.5cm}
\section*{Acknowledgements}
This manuscript was prepared by the BSN project (\url{https://bsnp.info/}).
We have made use of data from the European Space Agency (ESA) mission Gaia (\url{http://www.cosmos.esa.int/gaia}), processed by the Gaia Data Processing and Analysis Consortium (DPAC).
This work includes data from the TESS mission observations. Funding for the TESS mission is provided by the NASA Explorer Program.

\vspace{1.5cm}
\section*{ORCID iDs}
Atila Poro: 0000-0002-0196-9732\\
Sabrina Baudart: 0009-0004-8426-4114\\
Mahshid Nourmohammad: 0009-0006-0476-2055\\
Zahra Sabaghpour Arani: 0009-0009-7505-9190\\
Fatemeh Farhadi: 0009-0004-6762-9863\\
Selda Ranjbar Salehian: 0000-0002-5223-1332\\
Ahmad Sarostad: 0000-0001-6485-8696\\
Saeideh Ranjbaryan Iri Olya: 0009-0001-9087-2448\\
Maryam Hadizadeh: 0000-0003-1493-0295\\
AmirHossein Khodaei: 0009-0002-0058-415X\\

\vspace{1.5cm}
\section*{Appendix}
The appendix tables are related to the extracted TESS primary and secondary times of minima for the V1023 Her and V1397 Her binary systems.

\clearpage
\begin{table*}
\caption{Extracted times of minima for V1023 Her. All times of minima have been reduced to 2400000. The error value for the minimum times extracted from TESS is 0.0001.}
\centering
\changefontsizes{8}
\begin{center}
\renewcommand{\arraystretch}{0.84}
\footnotesize
\begin{tabular}{c c c c c c c c c c c c}
 \hline
 \hline
Min. & Epoch & O-C & Min. & Epoch & O-C & Min. & Epoch & O-C & Min. & Epoch & O-C\\
\hline
58929.7499	&	-2194.5	&	0.0051	&	58944.8956	&	-2147.5	&	0.0058	&	58957.4632	&	-2108.5	&	0.0063	&	58969.7077	&	-2070.5	&	0.0059	\\
58929.9136	&	-2194.0	&	0.0077	&	58945.0574	&	-2147.0	&	0.0065	&	58957.6242	&	-2108.0	&	0.0062	&	58969.8693	&	-2070.0	&	0.0064	\\
58930.0726	&	-2193.5	&	0.0056	&	58945.2181	&	-2146.5	&	0.0061	&	58957.7855	&	-2107.5	&	0.0063	&	58970.0301	&	-2069.5	&	0.0060	\\
58930.2357	&	-2193.0	&	0.0076	&	58945.3803	&	-2146.0	&	0.0072	&	58957.9463	&	-2107.0	&	0.0061	&	58970.1919	&	-2069.0	&	0.0068	\\
58930.3944	&	-2192.5	&	0.0051	&	58945.5398	&	-2145.5	&	0.0055	&	58958.1071	&	-2106.5	&	0.0057	&	58970.3522	&	-2068.5	&	0.0060	\\
58930.5577	&	-2192.0	&	0.0073	&	58945.7020	&	-2145.0	&	0.0066	&	58958.2691	&	-2106.0	&	0.0066	&	58970.5138	&	-2068.0	&	0.0064	\\
58930.7173	&	-2191.5	&	0.0059	&	58945.8625	&	-2144.5	&	0.0060	&	58958.4296	&	-2105.5	&	0.0059	&	58970.6746	&	-2067.5	&	0.0061	\\
58930.8803	&	-2191.0	&	0.0077	&	58946.0243	&	-2144.0	&	0.0067	&	58958.5910	&	-2105.0	&	0.0063	&	58970.9964	&	-2066.5	&	0.0056	\\
58931.0397	&	-2190.5	&	0.0060	&	58946.1842	&	-2143.5	&	0.0055	&	58958.7520	&	-2104.5	&	0.0061	&	58971.1584	&	-2066.0	&	0.0065	\\
58931.2020	&	-2190.0	&	0.0071	&	58946.3465	&	-2143.0	&	0.0067	&	58958.9131	&	-2104.0	&	0.0061	&	58971.3188	&	-2065.5	&	0.0059	\\
58931.3629	&	-2189.5	&	0.0069	&	58946.5070	&	-2142.5	&	0.0060	&	58959.0744	&	-2103.5	&	0.0063	&	58971.4799	&	-2065.0	&	0.0058	\\
58931.5246	&	-2189.0	&	0.0075	&	58946.6690	&	-2142.0	&	0.0070	&	58959.2352	&	-2103.0	&	0.0060	&	58971.6417	&	-2064.5	&	0.0065	\\
58931.6840	&	-2188.5	&	0.0058	&	58946.8284	&	-2141.5	&	0.0052	&	58959.3966	&	-2102.5	&	0.0063	&	58971.8027	&	-2064.0	&	0.0064	\\
58931.8467	&	-2188.0	&	0.0074	&	58946.9914	&	-2141.0	&	0.0071	&	58959.5575	&	-2102.0	&	0.0061	&	58971.9639	&	-2063.5	&	0.0064	\\
58932.0064	&	-2187.5	&	0.0060	&	58947.1505	&	-2140.5	&	0.0051	&	58959.7187	&	-2101.5	&	0.0062	&	58972.1251	&	-2063.0	&	0.0065	\\
58932.1692	&	-2187.0	&	0.0077	&	58947.3135	&	-2140.0	&	0.0069	&	58959.8796	&	-2101.0	&	0.0059	&	58972.2863	&	-2062.5	&	0.0066	\\
58932.3284	&	-2186.5	&	0.0057	&	58947.4728	&	-2139.5	&	0.0052	&	58960.0405	&	-2100.5	&	0.0057	&	58972.4472	&	-2062.0	&	0.0064	\\
58932.4914	&	-2186.0	&	0.0076	&	58947.6359	&	-2139.0	&	0.0072	&	58960.2026	&	-2100.0	&	0.0067	&	58972.6080	&	-2061.5	&	0.0060	\\
58932.6508	&	-2185.5	&	0.0059	&	58947.7953	&	-2138.5	&	0.0054	&	58960.3632	&	-2099.5	&	0.0061	&	58972.7694	&	-2061.0	&	0.0064	\\
58932.8136	&	-2185.0	&	0.0076	&	58947.9579	&	-2138.0	&	0.0069	&	58960.5246	&	-2099.0	&	0.0065	&	58972.9305	&	-2060.5	&	0.0063	\\
58932.9731	&	-2184.5	&	0.0059	&	58948.1174	&	-2137.5	&	0.0053	&	58960.6851	&	-2098.5	&	0.0059	&	58973.0918	&	-2060.0	&	0.0066	\\
58933.1355	&	-2184.0	&	0.0073	&	58948.2802	&	-2137.0	&	0.0070	&	58960.8468	&	-2098.0	&	0.0064	&	58973.2527	&	-2059.5	&	0.0064	\\
58933.2953	&	-2183.5	&	0.0059	&	58948.4397	&	-2136.5	&	0.0053	&	58961.0071	&	-2097.5	&	0.0056	&	58973.4140	&	-2059.0	&	0.0065	\\
58933.4581	&	-2183.0	&	0.0076	&	58948.6028	&	-2136.0	&	0.0073	&	58961.1688	&	-2097.0	&	0.0062	&	58973.5749	&	-2058.5	&	0.0063	\\
58933.6171	&	-2182.5	&	0.0056	&	58948.7622	&	-2135.5	&	0.0056	&	58961.3296	&	-2096.5	&	0.0059	&	58973.7366	&	-2058.0	&	0.0069	\\
58933.7806	&	-2182.0	&	0.0079	&	58948.9246	&	-2135.0	&	0.0068	&	58961.4913	&	-2096.0	&	0.0064	&	58973.8969	&	-2057.5	&	0.0061	\\
58933.9397	&	-2181.5	&	0.0058	&	58949.0847	&	-2134.5	&	0.0059	&	58961.6521	&	-2095.5	&	0.0061	&	58974.0583	&	-2057.0	&	0.0064	\\
58934.1028	&	-2181.0	&	0.0079	&	58949.2469	&	-2134.0	&	0.0070	&	58961.8135	&	-2095.0	&	0.0064	&	58974.2188	&	-2056.5	&	0.0057	\\
58934.2621	&	-2180.5	&	0.0060	&	58949.4067	&	-2133.5	&	0.0056	&	58961.9739	&	-2094.5	&	0.0057	&	58974.3809	&	-2056.0	&	0.0067	\\
58934.4243	&	-2180.0	&	0.0072	&	58949.5686	&	-2133.0	&	0.0064	&	58962.1355	&	-2094.0	&	0.0062	&	58974.5411	&	-2055.5	&	0.0058	\\
58934.5845	&	-2179.5	&	0.0063	&	58949.7295	&	-2132.5	&	0.0062	&	58962.2963	&	-2093.5	&	0.0059	&	58974.7030	&	-2055.0	&	0.0066	\\
58934.7467	&	-2179.0	&	0.0073	&	58949.8912	&	-2132.0	&	0.0068	&	58962.4576	&	-2093.0	&	0.0060	&	58974.8634	&	-2054.5	&	0.0059	\\
58934.9068	&	-2178.5	&	0.0062	&	58950.0514	&	-2131.5	&	0.0058	&	58962.6184	&	-2092.5	&	0.0057	&	58975.0253	&	-2054.0	&	0.0066	\\
58935.0688	&	-2178.0	&	0.0072	&	58950.2133	&	-2131.0	&	0.0067	&	58962.7799	&	-2092.0	&	0.0061	&	58975.1854	&	-2053.5	&	0.0056	\\
58935.5514	&	-2176.5	&	0.0064	&	58950.3735	&	-2130.5	&	0.0057	&	58962.9409	&	-2091.5	&	0.0060	&	58975.3476	&	-2053.0	&	0.0067	\\
58935.7133	&	-2176.0	&	0.0072	&	58950.5350	&	-2130.0	&	0.0061	&	58963.1025	&	-2091.0	&	0.0064	&	58975.5086	&	-2052.5	&	0.0066	\\
58935.8730	&	-2175.5	&	0.0057	&	58950.6962	&	-2129.5	&	0.0062	&	58963.2625	&	-2090.5	&	0.0054	&	58975.6705	&	-2052.0	&	0.0073	\\
58936.0356	&	-2175.0	&	0.0073	&	58950.8576	&	-2129.0	&	0.0065	&	58963.4244	&	-2090.0	&	0.0062	&	58975.8304	&	-2051.5	&	0.0061	\\
58936.1955	&	-2174.5	&	0.0060	&	58951.0177	&	-2128.5	&	0.0055	&	58963.5852	&	-2089.5	&	0.0058	&	58975.9921	&	-2051.0	&	0.0068	\\
58936.3580	&	-2174.0	&	0.0074	&	58951.1797	&	-2128.0	&	0.0064	&	58963.7469	&	-2089.0	&	0.0064	&	58976.1527	&	-2050.5	&	0.0062	\\
58936.5177	&	-2173.5	&	0.0060	&	58951.3405	&	-2127.5	&	0.0060	&	58963.9077	&	-2088.5	&	0.0061	&	58976.3141	&	-2050.0	&	0.0065	\\
58936.6801	&	-2173.0	&	0.0073	&	58951.5024	&	-2127.0	&	0.0068	&	58964.0680	&	-2088.0	&	0.0053	&	58976.4748	&	-2049.5	&	0.0061	\\
58936.8399	&	-2172.5	&	0.0059	&	58951.6625	&	-2126.5	&	0.0058	&	58964.2296	&	-2087.5	&	0.0057	&	58976.6366	&	-2049.0	&	0.0068	\\
58937.0022	&	-2172.0	&	0.0072	&	58951.8243	&	-2126.0	&	0.0065	&	58964.3912	&	-2087.0	&	0.0062	&	58976.7968	&	-2048.5	&	0.0058	\\
58937.1623	&	-2171.5	&	0.0061	&	58951.9850	&	-2125.5	&	0.0060	&	58964.5518	&	-2086.5	&	0.0057	&	58976.9591	&	-2048.0	&	0.0070	\\
58937.3245	&	-2171.0	&	0.0072	&	58952.1469	&	-2125.0	&	0.0068	&	58964.7131	&	-2086.0	&	0.0059	&	58977.1190	&	-2047.5	&	0.0058	\\
58937.4839	&	-2170.5	&	0.0055	&	58952.3068	&	-2124.5	&	0.0056	&	58964.8747	&	-2085.5	&	0.0064	&	58977.2812	&	-2047.0	&	0.0069	\\
58937.6464	&	-2170.0	&	0.0068	&	58952.4686	&	-2124.0	&	0.0064	&	58965.0355	&	-2085.0	&	0.0061	&	58977.4409	&	-2046.5	&	0.0055	\\
58937.8071	&	-2169.5	&	0.0064	&	58952.6293	&	-2123.5	&	0.0059	&	58965.1964	&	-2084.5	&	0.0058	&	58977.6032	&	-2046.0	&	0.0067	\\
58937.9691	&	-2169.0	&	0.0073	&	58952.7910	&	-2123.0	&	0.0065	&	58965.3576	&	-2084.0	&	0.0059	&	58977.7636	&	-2045.5	&	0.0060	\\
58938.1287	&	-2168.5	&	0.0058	&	58952.9514	&	-2122.5	&	0.0058	&	58965.5183	&	-2083.5	&	0.0055	&	58977.9258	&	-2045.0	&	0.0070	\\
58938.2913	&	-2168.0	&	0.0073	&	58953.1130	&	-2122.0	&	0.0062	&	58965.6803	&	-2083.0	&	0.0064	&	58978.0856	&	-2044.5	&	0.0057	\\
58938.4510	&	-2167.5	&	0.0059	&	58953.2739	&	-2121.5	&	0.0060	&	58965.8406	&	-2082.5	&	0.0056	&	58978.2479	&	-2044.0	&	0.0069	\\
58938.6136	&	-2167.0	&	0.0074	&	58953.4355	&	-2121.0	&	0.0065	&	58966.0030	&	-2082.0	&	0.0069	&	58978.4074	&	-2043.5	&	0.0053	\\
58938.7734	&	-2166.5	&	0.0061	&	58953.5960	&	-2120.5	&	0.0059	&	58966.1624	&	-2081.5	&	0.0052	&	58978.5707	&	-2043.0	&	0.0074	\\
58938.9355	&	-2166.0	&	0.0071	&	58953.7574	&	-2120.0	&	0.0062	&	58966.3249	&	-2081.0	&	0.0065	&	58978.7300	&	-2042.5	&	0.0057	\\
58939.0951	&	-2165.5	&	0.0055	&	58953.9185	&	-2119.5	&	0.0062	&	58966.4848	&	-2080.5	&	0.0053	&	58978.8920	&	-2042.0	&	0.0065	\\
58939.2576	&	-2165.0	&	0.0069	&	58954.0797	&	-2119.0	&	0.0062	&	58966.6475	&	-2080.0	&	0.0069	&	58979.0523	&	-2041.5	&	0.0057	\\
58939.4173	&	-2164.5	&	0.0055	&	58954.2405	&	-2118.5	&	0.0059	&	58966.8076	&	-2079.5	&	0.0059	&	58979.2144	&	-2041.0	&	0.0067	\\
58939.5804	&	-2164.0	&	0.0074	&	58954.4023	&	-2118.0	&	0.0067	&	58966.9693	&	-2079.0	&	0.0065	&	58979.3746	&	-2040.5	&	0.0058	\\
58939.7398	&	-2163.5	&	0.0058	&	58954.5629	&	-2117.5	&	0.0060	&	58967.1291	&	-2078.5	&	0.0051	&	58979.5367	&	-2040.0	&	0.0068	\\
58939.9025	&	-2163.0	&	0.0074	&	58954.7245	&	-2117.0	&	0.0066	&	58967.2916	&	-2078.0	&	0.0066	&	58979.6969	&	-2039.5	&	0.0059	\\
58940.0617	&	-2162.5	&	0.0055	&	58955.8522	&	-2113.5	&	0.0065	&	58967.4527	&	-2077.5	&	0.0065	&	58979.8588	&	-2039.0	&	0.0066	\\
58940.2246	&	-2162.0	&	0.0072	&	58956.0127	&	-2113.0	&	0.0059	&	58967.6133	&	-2077.0	&	0.0060	&	58980.0184	&	-2038.5	&	0.0051	\\
58940.3841	&	-2161.5	&	0.0056	&	58956.1748	&	-2112.5	&	0.0068	&	58967.7740	&	-2076.5	&	0.0056	&	58980.1816	&	-2038.0	&	0.0072	\\
58940.5466	&	-2161.0	&	0.0070	&	58956.3355	&	-2112.0	&	0.0064	&	58967.9357	&	-2076.0	&	0.0062	&	58980.3409	&	-2037.5	&	0.0053	\\
58940.7069	&	-2160.5	&	0.0062	&	58956.4970	&	-2111.5	&	0.0068	&	58968.0972	&	-2075.5	&	0.0065	&	58980.5035	&	-2037.0	&	0.0068	\\
58944.4140	&	-2149.0	&	0.0075	&	58956.6578	&	-2111.0	&	0.0065	&	58968.2581	&	-2075.0	&	0.0064	&	58980.6635	&	-2036.5	&	0.0057	\\
58944.5733	&	-2148.5	&	0.0058	&	58956.8197	&	-2110.5	&	0.0072	&	58969.3854	&	-2071.5	&	0.0058	&	58980.8256	&	-2036.0	&	0.0068	\\
58944.7352	&	-2148.0	&	0.0066	&	58957.3020	&	-2109.0	&	0.0062	&	58969.5470	&	-2071.0	&	0.0063	&	58980.9855	&	-2035.5	&	0.0055	\\
\hline
\hline
\end{tabular}
\end{center}
\label{tabA1}
\end{table*}

\begin{table*}
\caption{Continued for V1023 Her.}
\centering
\begin{center}
\changefontsizes{8}
\footnotesize
\begin{tabular}{c c c c c c c c c c c c}
 \hline
 \hline
Min. & Epoch & O-C & Min. & Epoch & O-C & Min. & Epoch & O-C & Min. & Epoch & O-C\\
\hline
58981.1479	&	-2035.0	&	0.0068	&	59677.3351	&	125.5	&	0.0072	&	59691.3486	&	169.0	&	0.0035	&	59706.1715	&	215.0	&	0.0037	\\
58981.3076	&	-2034.5	&	0.0054	&	59677.4930	&	126.0	&	0.0040	&	59693.1246	&	174.5	&	0.0072	&	59706.3371	&	215.5	&	0.0081	\\
58981.4704	&	-2034.0	&	0.0071	&	59677.6573	&	126.5	&	0.0072	&	59693.2824	&	175.0	&	0.0039	&	59706.4937	&	216.0	&	0.0036	\\
58981.6294	&	-2033.5	&	0.0049	&	59677.8156	&	127.0	&	0.0044	&	59693.4468	&	175.5	&	0.0072	&	59706.6590	&	216.5	&	0.0078	\\
58981.7926	&	-2033.0	&	0.0071	&	59677.9795	&	127.5	&	0.0071	&	59693.6049	&	176.0	&	0.0042	&	59706.8159	&	217.0	&	0.0036	\\
58981.9516	&	-2032.5	&	0.0049	&	59678.1377	&	128.0	&	0.0042	&	59696.9916	&	186.5	&	0.0075	&	59706.9812	&	217.5	&	0.0078	\\
58982.1145	&	-2032.0	&	0.0066	&	59678.3021	&	128.5	&	0.0075	&	59697.1495	&	187.0	&	0.0042	&	59710.0380	&	227.0	&	0.0033	\\
59665.4117	&	88.5	&	0.0065	&	59683.4580	&	144.5	&	0.0076	&	59697.3137	&	187.5	&	0.0073	&	59710.2042	&	227.5	&	0.0084	\\
59665.5696	&	89.0	&	0.0032	&	59683.6159	&	145.0	&	0.0045	&	59697.4712	&	188.0	&	0.0037	&	59710.3601	&	228.0	&	0.0032	\\
59669.6010	&	101.5	&	0.0067	&	59683.7794	&	145.5	&	0.0068	&	59697.6361	&	188.5	&	0.0075	&	59710.5258	&	228.5	&	0.0078	\\
59669.7583	&	102.0	&	0.0029	&	59683.9376	&	146.0	&	0.0039	&	59697.7938	&	189.0	&	0.0040	&	59710.6819	&	229.0	&	0.0028	\\
59669.9238	&	102.5	&	0.0073	&	59684.1022	&	146.5	&	0.0074	&	59697.9583	&	189.5	&	0.0074	&	59710.8486	&	229.5	&	0.0084	\\
59670.0807	&	103.0	&	0.0031	&	59684.2594	&	147.0	&	0.0035	&	59698.1158	&	190.0	&	0.0038	&	59711.0044	&	230.0	&	0.0031	\\
59670.2454	&	103.5	&	0.0066	&	59684.4243	&	147.5	&	0.0073	&	59698.2804	&	190.5	&	0.0073	&	59711.1707	&	230.5	&	0.0082	\\
59670.4036	&	104.0	&	0.0037	&	59684.5821	&	148.0	&	0.0039	&	59698.4379	&	191.0	&	0.0036	&	59711.3272	&	231.0	&	0.0036	\\
59670.5681	&	104.5	&	0.0071	&	59684.7464	&	148.5	&	0.0072	&	59698.6028	&	191.5	&	0.0075	&	59711.4928	&	231.5	&	0.0081	\\
59670.7251	&	105.0	&	0.0030	&	59684.9042	&	149.0	&	0.0038	&	59698.7596	&	192.0	&	0.0032	&	59711.6489	&	232.0	&	0.0030	\\
59670.8896	&	105.5	&	0.0064	&	59685.0690	&	149.5	&	0.0075	&	59698.9248	&	192.5	&	0.0073	&	59711.8150	&	232.5	&	0.0080	\\
59671.0478	&	106.0	&	0.0035	&	59685.2265	&	150.0	&	0.0038	&	59699.0824	&	193.0	&	0.0037	&	59711.9713	&	233.0	&	0.0032	\\
59671.2124	&	106.5	&	0.0070	&	59685.3906	&	150.5	&	0.0068	&	59699.2473	&	193.5	&	0.0074	&	59712.1376	&	233.5	&	0.0084	\\
59671.3700	&	107.0	&	0.0034	&	59685.5490	&	151.0	&	0.0041	&	59699.4044	&	194.0	&	0.0035	&	59712.2934	&	234.0	&	0.0031	\\
59671.5338	&	107.5	&	0.0061	&	59685.7132	&	151.5	&	0.0072	&	59699.5693	&	194.5	&	0.0073	&	59712.4592	&	234.5	&	0.0078	\\
59671.6922	&	108.0	&	0.0034	&	59685.8713	&	152.0	&	0.0042	&	59699.7266	&	195.0	&	0.0034	&	59712.6159	&	235.0	&	0.0034	\\
59671.8567	&	108.5	&	0.0068	&	59686.0356	&	152.5	&	0.0074	&	59699.8913	&	195.5	&	0.0070	&	59712.7815	&	235.5	&	0.0078	\\
59672.0144	&	109.0	&	0.0034	&	59686.1934	&	153.0	&	0.0041	&	59700.0490	&	196.0	&	0.0036	&	59712.9380	&	236.0	&	0.0032	\\
59672.1789	&	109.5	&	0.0068	&	59686.3574	&	153.5	&	0.0070	&	59700.2139	&	196.5	&	0.0073	&	59713.1040	&	236.5	&	0.0081	\\
59672.5010	&	110.5	&	0.0066	&	59686.5158	&	154.0	&	0.0042	&	59700.3709	&	197.0	&	0.0033	&	59713.2601	&	237.0	&	0.0031	\\
59672.6591	&	111.0	&	0.0036	&	59686.6797	&	154.5	&	0.0070	&	59700.5357	&	197.5	&	0.0070	&	59713.4259	&	237.5	&	0.0077	\\
59672.8232	&	111.5	&	0.0066	&	59686.8378	&	155.0	&	0.0040	&	59700.6932	&	198.0	&	0.0033	&	59713.5826	&	238.0	&	0.0033	\\
59672.9812	&	112.0	&	0.0035	&	59687.0021	&	155.5	&	0.0072	&	59700.8586	&	198.5	&	0.0076	&	59713.7479	&	238.5	&	0.0075	\\
59673.1458	&	112.5	&	0.0069	&	59687.1601	&	156.0	&	0.0041	&	59701.0159	&	199.0	&	0.0038	&	59713.9046	&	239.0	&	0.0031	\\
59673.3034	&	113.0	&	0.0035	&	59687.3245	&	156.5	&	0.0074	&	59701.1807	&	199.5	&	0.0075	&	59714.0703	&	239.5	&	0.0078	\\
59673.4683	&	113.5	&	0.0072	&	59687.4824	&	157.0	&	0.0041	&	59701.3376	&	200.0	&	0.0033	&	59714.2265	&	240.0	&	0.0028	\\
59673.6257	&	114.0	&	0.0035	&	59687.6464	&	157.5	&	0.0070	&	59701.5027	&	200.5	&	0.0073	&	59714.3929	&	240.5	&	0.0081	\\
59673.7901	&	114.5	&	0.0068	&	59687.8048	&	158.0	&	0.0043	&	59701.6601	&	201.0	&	0.0035	&	59714.5490	&	241.0	&	0.0031	\\
59673.9480	&	115.0	&	0.0035	&	59687.9689	&	158.5	&	0.0072	&	59701.8249	&	201.5	&	0.0072	&	59714.7150	&	241.5	&	0.0079	\\
59674.1125	&	115.5	&	0.0069	&	59688.1271	&	159.0	&	0.0043	&	59701.9823	&	202.0	&	0.0035	&	59714.8710	&	242.0	&	0.0028	\\
59674.2701	&	116.0	&	0.0034	&	59688.2904	&	159.5	&	0.0065	&	59702.1471	&	202.5	&	0.0071	&	59715.0370	&	242.5	&	0.0077	\\
59674.4348	&	116.5	&	0.0070	&	59688.4492	&	160.0	&	0.0043	&	59702.3049	&	203.0	&	0.0039	&	59715.1932	&	243.0	&	0.0028	\\
59674.5924	&	117.0	&	0.0035	&	59688.6128	&	160.5	&	0.0067	&	59702.4697	&	203.5	&	0.0075	&	59715.3593	&	243.5	&	0.0078	\\
59674.7568	&	117.5	&	0.0068	&	59688.7714	&	161.0	&	0.0042	&	59702.6272	&	204.0	&	0.0040	&	59715.5157	&	244.0	&	0.0031	\\
59674.9148	&	118.0	&	0.0037	&	59688.9354	&	161.5	&	0.0070	&	59702.7913	&	204.5	&	0.0070	&	59715.6816	&	244.5	&	0.0078	\\
59675.0791	&	118.5	&	0.0068	&	59689.0938	&	162.0	&	0.0043	&	59702.9494	&	205.0	&	0.0039	&	59715.8381	&	245.0	&	0.0032	\\
59675.2373	&	119.0	&	0.0039	&	59689.2573	&	162.5	&	0.0067	&	59703.1141	&	205.5	&	0.0075	&	59716.0038	&	245.5	&	0.0078	\\
59675.4017	&	119.5	&	0.0072	&	59689.4155	&	163.0	&	0.0038	&	59703.2713	&	206.0	&	0.0035	&	59716.1600	&	246.0	&	0.0029	\\
59675.5594	&	120.0	&	0.0037	&	59689.5799	&	163.5	&	0.0071	&	59703.4364	&	206.5	&	0.0075	&	59716.3260	&	246.5	&	0.0078	\\
59675.7233	&	120.5	&	0.0065	&	59689.7380	&	164.0	&	0.0041	&	59703.5936	&	207.0	&	0.0037	&	59716.4823	&	247.0	&	0.0030	\\
59675.8823	&	121.0	&	0.0045	&	59689.9020	&	164.5	&	0.0069	&	59703.7583	&	207.5	&	0.0072	&	59716.6481	&	247.5	&	0.0077	\\
59676.0458	&	121.5	&	0.0069	&	59690.0601	&	165.0	&	0.0040	&	59703.9156	&	208.0	&	0.0034	&	59716.8048	&	248.0	&	0.0032	\\
59676.2043	&	122.0	&	0.0042	&	59690.2236	&	165.5	&	0.0063	&	59704.0808	&	208.5	&	0.0075	&	59716.9704	&	248.5	&	0.0077	\\
59676.3681	&	122.5	&	0.0069	&	59690.3825	&	166.0	&	0.0041	&	59704.2380	&	209.0	&	0.0035	&	59717.1271	&	249.0	&	0.0032	\\
59676.5262	&	123.0	&	0.0039	&	59690.5464	&	166.5	&	0.0069	&	59704.4030	&	209.5	&	0.0074	&	59717.2929	&	249.5	&	0.0080	\\
59676.6901	&	123.5	&	0.0067	&	59690.7044	&	167.0	&	0.0038	&	59704.5603	&	210.0	&	0.0036	&	59717.4491	&	250.0	&	0.0031	\\
59676.8488	&	124.0	&	0.0043	&	59690.8688	&	167.5	&	0.0071	&	59705.7282	&	213.5	&	0.0437	&		&		&		\\
59677.0126	&	124.5	&	0.0069	&	59691.0268	&	168.0	&	0.0039	&	59705.8492	&	214.0	&	0.0036	&		&		&		\\
59677.1714	&	125.0	&	0.0046	&	59691.1911	&	168.5	&	0.0071	&	59706.0149	&	214.5	&	0.0081	&		&		&		\\
\hline
\hline
\end{tabular}
\end{center}
\label{tabA2}
\end{table*}

\begin{table*}
\caption{Available times of minima for V1397 Her. All times of minima have been reduced to 2400000. The error value for the minimum times extracted from TESS is 0.0001.}
\centering
\changefontsizes{8}
\begin{center}
\footnotesize
\begin{tabular}{c c c c c c c c c c c c}
 \hline
 \hline
Min. & Epoch & O-C & Min. & Epoch & O-C & Min. & Epoch & O-C & Min. & Epoch & O-C\\
\hline
58983.7128	&	13282	&	0.0439	&	58993.2124	&	13306.5	&	0.0438	&	59004.0692	&	13334.5	&	0.0437	&	59728.1921	&	15202	&	0.0528	\\
58983.9061	&	13282.5	&	0.0434	&	58993.4064	&	13307	&	0.0440	&	59004.2635	&	13335	&	0.0441	&	59728.3864	&	15202.5	&	0.0533	\\
58984.1005	&	13283	&	0.0439	&	58993.5999	&	13307.5	&	0.0435	&	59004.4570	&	13335.5	&	0.0438	&	59728.5798	&	15203	&	0.0528	\\
58984.2939	&	13283.5	&	0.0435	&	58993.7942	&	13308	&	0.0440	&	59004.6511	&	13336	&	0.0440	&	59728.7736	&	15203.5	&	0.0528	\\
58984.4882	&	13284	&	0.0439	&	58993.9877	&	13308.5	&	0.0437	&	59004.8446	&	13336.5	&	0.0437	&	59728.9683	&	15204	&	0.0536	\\
58984.6814	&	13284.5	&	0.0432	&	58994.1820	&	13309	&	0.0440	&	59005.0455	&	13337	&	0.0507	&	59729.1614	&	15204.5	&	0.0529	\\
58984.8760	&	13285	&	0.0439	&	58994.3758	&	13309.5	&	0.0440	&	59005.2323	&	13337.5	&	0.0436	&	59729.3554	&	15205	&	0.0529	\\
58985.0696	&	13285.5	&	0.0437	&	58994.5697	&	13310	&	0.0440	&	59005.4268	&	13338	&	0.0443	&	59729.5488	&	15205.5	&	0.0524	\\
58985.2636	&	13286	&	0.0438	&	58994.7634	&	13310.5	&	0.0438	&	59005.6201	&	13338.5	&	0.0437	&	59729.7431	&	15206	&	0.0529	\\
58985.4572	&	13286.5	&	0.0436	&	58994.9574	&	13311	&	0.0440	&	59005.8144	&	13339	&	0.0441	&	59729.9369	&	15206.5	&	0.0528	\\
58985.6512	&	13287	&	0.0437	&	58995.1510	&	13311.5	&	0.0437	&	59006.0087	&	13339.5	&	0.0446	&	59730.1309	&	15207	&	0.0530	\\
58985.8449	&	13287.5	&	0.0435	&	58995.3452	&	13312	&	0.0440	&	59006.2022	&	13340	&	0.0442	&	59731.4879	&	15210.5	&	0.0529	\\
58986.0393	&	13288	&	0.0440	&	58995.5389	&	13312.5	&	0.0439	&	59006.3957	&	13340.5	&	0.0437	&	59731.6820	&	15211	&	0.0531	\\
58986.2329	&	13288.5	&	0.0437	&	58997.0899	&	13316.5	&	0.0439	&	59006.5900	&	13341	&	0.0442	&	59731.8756	&	15211.5	&	0.0528	\\
58986.4269	&	13289	&	0.0439	&	58997.2841	&	13317	&	0.0442	&	59006.7834	&	13341.5	&	0.0437	&	59732.0652	&	15212	&	0.0485	\\
58986.6204	&	13289.5	&	0.0435	&	58997.4777	&	13317.5	&	0.0439	&	59006.9777	&	13342	&	0.0441	&	59732.2633	&	15212.5	&	0.0528	\\
58986.8143	&	13290	&	0.0436	&	58997.6717	&	13318	&	0.0441	&	59007.1712	&	13342.5	&	0.0438	&	59732.4575	&	15213	&	0.0531	\\
58987.0082	&	13290.5	&	0.0435	&	58997.8654	&	13318.5	&	0.0439	&	59007.3655	&	13343	&	0.0442	&	59732.6513	&	15213.5	&	0.0530	\\
58987.2023	&	13291	&	0.0438	&	58998.0596	&	13319	&	0.0442	&	59007.5588	&	13343.5	&	0.0436	&	59732.8452	&	15214	&	0.0530	\\
58987.3960	&	13291.5	&	0.0436	&	58998.2524	&	13319.5	&	0.0431	&	59007.7532	&	13344	&	0.0442	&	59737.4979	&	15226	&	0.0528	\\
58987.5902	&	13292	&	0.0439	&	58998.4472	&	13320	&	0.0440	&	59007.9467	&	13344.5	&	0.0438	&	59737.6917	&	15226.5	&	0.0528	\\
58987.7838	&	13292.5	&	0.0437	&	58998.6408	&	13320.5	&	0.0438	&	59008.1410	&	13345	&	0.0442	&	59737.8858	&	15227	&	0.0530	\\
58987.9779	&	13293	&	0.0438	&	58998.8351	&	13321	&	0.0442	&	59008.3345	&	13345.5	&	0.0439	&	59738.0794	&	15227.5	&	0.0527	\\
58988.1718	&	13293.5	&	0.0440	&	58999.0280	&	13321.5	&	0.0433	&	59008.5286	&	13346	&	0.0441	&	59738.2735	&	15228	&	0.0529	\\
58988.3655	&	13294	&	0.0437	&	58999.2227	&	13322	&	0.0441	&	59008.7220	&	13346.5	&	0.0436	&	59738.4671	&	15228.5	&	0.0527	\\
58988.5588	&	13294.5	&	0.0432	&	58999.4163	&	13322.5	&	0.0438	&	59008.9165	&	13347	&	0.0443	&	59738.6613	&	15229	&	0.0530	\\
58988.7533	&	13295	&	0.0438	&	58999.6106	&	13323	&	0.0442	&	59718.6922	&	15177.5	&	0.0527	&	59738.8547	&	15229.5	&	0.0525	\\
58988.9471	&	13295.5	&	0.0437	&	58999.8040	&	13323.5	&	0.0438	&	59718.8861	&	15178	&	0.0528	&	59739.0490	&	15230	&	0.0529	\\
58989.1411	&	13296	&	0.0438	&	58999.9982	&	13324	&	0.0441	&	59719.0800	&	15178.5	&	0.0527	&	59739.2434	&	15230.5	&	0.0535	\\
58989.3348	&	13296.5	&	0.0437	&	59000.1917	&	13324.5	&	0.0437	&	59719.2739	&	15179	&	0.0528	&	59739.4366	&	15231	&	0.0528	\\
58989.5287	&	13297	&	0.0437	&	59000.3860	&	13325	&	0.0442	&	59719.4676	&	15179.5	&	0.0526	&	59739.6304	&	15231.5	&	0.0527	\\
58989.7227	&	13297.5	&	0.0438	&	59000.5795	&	13325.5	&	0.0438	&	59719.6617	&	15180	&	0.0528	&	59739.8245	&	15232	&	0.0529	\\
58989.9166	&	13298	&	0.0438	&	59000.7738	&	13326	&	0.0442	&	59724.8961	&	15193.5	&	0.0527	&	59740.0085	&	15232.5	&	0.0430	\\
58990.1105	&	13298.5	&	0.0439	&	59000.9674	&	13326.5	&	0.0439	&	59725.0901	&	15194	&	0.0528	&	59740.2123	&	15233	&	0.0530	\\
58990.3042	&	13299	&	0.0437	&	59001.1616	&	13327	&	0.0442	&	59725.2838	&	15194.5	&	0.0527	&	59740.4060	&	15233.5	&	0.0528	\\
58990.4982	&	13299.5	&	0.0438	&	59001.3550	&	13327.5	&	0.0438	&	59725.4778	&	15195	&	0.0528	&	59740.6000	&	15234	&	0.0530	\\
58990.6920	&	13300	&	0.0438	&	59001.5491	&	13328	&	0.0440	&	59725.6716	&	15195.5	&	0.0527	&	59740.7938	&	15234.5	&	0.0528	\\
58990.8859	&	13300.5	&	0.0438	&	59001.7427	&	13328.5	&	0.0437	&	59725.8655	&	15196	&	0.0528	&	59740.9878	&	15235	&	0.0530	\\
58991.0798	&	13301	&	0.0438	&	59001.9370	&	13329	&	0.0442	&	59726.0598	&	15196.5	&	0.0531	&	59741.1815	&	15235.5	&	0.0528	\\
58991.2736	&	13301.5	&	0.0438	&	59002.1306	&	13329.5	&	0.0439	&	59726.2532	&	15197	&	0.0527	&	59741.3755	&	15236	&	0.0530	\\
58991.4676	&	13302	&	0.0438	&	59002.3246	&	13330	&	0.0440	&	59726.4471	&	15197.5	&	0.0528	&	59741.5693	&	15236.5	&	0.0529	\\
58991.6614	&	13302.5	&	0.0438	&	59002.5184	&	13330.5	&	0.0439	&	59726.6410	&	15198	&	0.0528	&	59741.7632	&	15237	&	0.0529	\\
58991.8552	&	13303	&	0.0437	&	59002.7123	&	13331	&	0.0440	&	59726.8348	&	15198.5	&	0.0526	&	59741.9570	&	15237.5	&	0.0528	\\
58992.0493	&	13303.5	&	0.0440	&	59002.9059	&	13331.5	&	0.0437	&	59727.0289	&	15199	&	0.0529	&	59742.1510	&	15238	&	0.0529	\\
58992.2432	&	13304	&	0.0440	&	59003.1003	&	13332	&	0.0443	&	59727.2219	&	15199.5	&	0.0520	&	59742.3441	&	15238.5	&	0.0522	\\
58992.4369	&	13304.5	&	0.0438	&	59003.2936	&	13332.5	&	0.0436	&	59727.4164	&	15200	&	0.0526	&	59742.5387	&	15239	&	0.0529	\\
58992.6307	&	13305	&	0.0437	&	59003.4879	&	13333	&	0.0441	&	59727.6102	&	15200.5	&	0.0526	&	59742.7326	&	15239.5	&	0.0529	\\
58992.8245	&	13305.5	&	0.0437	&	59003.6814	&	13333.5	&	0.0437	&	59727.8047	&	15201	&	0.0533	&	59742.9265	&	15240	&	0.0530	\\
58993.0185	&	13306	&	0.0438	&	59003.8756	&	13334	&	0.0440	&	59727.9981	&	15201.5	&	0.0528	&		&		&		\\
\hline
\hline
\end{tabular}
\end{center}
\label{tabA3}
\end{table*}

\clearpage
\bibliography{Ref}{}
\bibliographystyle{aasjournal}

\end{document}